\documentstyle[aps,prb,multicol,epsf]{revtex}  
\newcommand{\beq}{\begin{equation}}
\newcommand{\eeq}{\end{equation}}
\newcommand{\beqa}{\begin{eqnarray}}
\newcommand{\eeqa}{\end{eqnarray}}
\newcommand{\w}{\omega}

\newcommand{\nn}{{\vec n}}

\newcommand{\q}{{\vec q}}

\renewcommand{\>}{\rangle}
\newcommand{\<}{\langle}
\newcommand{\St}{\mbox{St}}
\renewcommand{\Re}{\mbox{Re} }
\renewcommand{\Im}{\mbox{Im} }
\begin{document}          
\title{Interaction corrections to the Hall coefficient
 at intermediate temperatures }
\author{G\'abor Zala$^{(a,b)}$, B.N. Narozhny$^{(a)}$, and I.L. Aleiner$^{(a,b)}$} 
\address{
$^{(a)}$Department of Physics and Astronomy, SUNY at Stony Brook, 
Stony Brook, NY, 11794, USA \\
$^{(b)}$Physics Department, Lancaster University, LA1 4YB, Lancaster, UK 
\\
} 
\maketitle
\begin{abstract}                

We investigate the effect of electron-electron interaction on the
temperature dependence of the Hall coefficient of 2D electron gas at
arbitrary relation between the temperature $T$ and the elastic
mean-free time $\tau$.  At small temperature $T\tau \ll \hbar$ we
reproduce the known relation between the logarithmic temperature
dependences of the Hall coefficient and of the longitudinal
conductivity. At higher temperatures, this relation is violated quite
rapidly; correction to the Hall coefficient becomes $\propto 1/T$
whereas the longitudinal conductivity becomes linear in temperature.
\end{abstract}

\begin{multicols}{2}

{\em Introduction} -- It is well known\cite{aar} that 
electron-electron interaction at low temperatures ($T\tau \ll \hbar$)
leads to the logarithmically divergent correction $\delta\sigma_{xx}$
to the longitudinal conductivity of 2D electron gas. Whereas for a
wide range of temperatures the sign and the magnitude of those
corrections are not universal (in particular due to the Fermi liquid
type interactions in the triplet channel), one property remains
intact\cite{halls}: there is no logarithmic interaction correction to
the Hall conductivity $\sigma_{xy}$ for any type of the interaction.
When the interaction correction is still smaller than the Drude
conductivity, the latter fact can be also re-written as 
\beq 
\gamma=2,
\label{two}
\eeq
where the parameter $\gamma$ is defined as
\beq
\gamma(T) = - 
\frac{\partial_T \ln \ \rho_{xy}(T)}{\partial_T \ln \sigma_{xx}(T) },
\label{ourgamma}
\eeq
with $\rho_{xy}$ being the Hall resistivity.

Equation (\ref{two}) is not controlled by any symmetries of the
system and comes about as a result of the diffusive approximation
justified only for $T\tau \ll \hbar$.  For the 2D electron gas based
on semiconductor heterostructures \cite{Kravchenko,Savchenko} 
this condition is easily violated. Therefore, one can no longer 
rely on the relation
(\ref{two}) for $T\tau \gtrsim \hbar$. The theory describing the
corresponding temperature behavior of $\gamma$ is still not
available. Our goal in this paper is to derive the corresponding
analytic formulas. Temperature dependence of the Hall coefficient
much weaker than predicted by Eq.~(\ref{ourgamma}) was recently
reported in Ref.~\cite{Savchenko}. 

{\em Method} -- We will use the kinetic equation formalism for the
interaction corrections which we briefly summarize below and then
apply it to the calculation of the Hall coefficient. The detailed
derivation of the formalism and the region of its validity can be
found in Ref.~\onlinecite{intermed}.

The electric current is expressed in terms of the distribution
function $f(t; \epsilon,\vec r,\vec n)$ as
\beq
{\vec J}(t,\vec r) =e\nu v_F \int\limits_{-\infty}^{\infty} d\epsilon
\langle \vec n f(t; \epsilon , \vec r ,\vec n)\rangle_n.
\label{current}
\eeq  
Here $\nu$ is the density of states of the noninteracting
system taken at the Fermi surface and $v_F$ is the Fermi velocity,
$\vec n = (\cos\theta, \sin\theta)$ is the unit vector in the direction 
of the electron momentum and angular averaging is 
$
\langle \dots \rangle_n =
\int 
\frac{d\theta}{2\pi} \dots
$.
The Boltzmann-like equation for the stationary and homogeneous 
distribution function is
\beq
\left[  
e v_F (\vec n {\vec E})\frac{\partial}{\partial \epsilon}
+\omega_c   
\vec{b}\left(\vec n \times \frac{\partial}{\partial \vec n} \right) 
\right] f =
\St\left\{f\right\}.
\label{Boltzmann}
\eeq
\noindent
where ${\vec E}$ is the applied electric field and $\omega_c$ is the
cyclotron frequency due to the external magnetic field and $\vec{b}$
is the unit vector along the field.

All of the interaction effects are taken into account~\cite{intermed}
in the elastic and inelastic parts of the collision integral, $
\St\left\{f\right\} = \St_{el}\left\{f\right\} +
\St_{in}\left\{f\right\} $.  The inelastic part does not contribute to
the Hall current.  The relevant, elastic part can be written as
(in all intermediate formulae we use the units with 
$\hbar = 1$)\beqa
&&\St_{el}\left\{f\right\}= - \frac{f(\epsilon,\vec n) -\langle
f(\epsilon,\vec n)\rangle_n}{\tau} \nonumber\\ && +I_0(\epsilon,\vec
n) \langle f( \epsilon,\vec n)\rangle_n
\label{elasticpart}
+n_\alpha I_1^{\alpha\beta} (\epsilon) \langle n_\beta 
f(\epsilon,\vec n)\rangle_n,  
\nonumber 
\eeqa 
where the second line describes the effect of interaction on elastic
collisions and 
\begin{mathletters}\label{I's}
\beqa
&&I_0(\epsilon,\vec n)= \label{I_0} 
- \frac{8}{\tau}  
\int \frac{d\omega}{2\pi}
\Big\{n_\alpha K_0^{\alpha\beta}(\omega)  
\langle n_\beta f( \epsilon-\omega,\vec n)\rangle_n
 \\
&&\quad
+\frac{n_\alpha L_0^{\alpha\beta}(\w )}{2} 
 e E_\beta\frac{\partial}{\partial \epsilon} 
\langle f(\epsilon-\omega,\vec n)\rangle_n
\Big\} \nonumber \\
&& I_1^{\alpha\beta}(\epsilon)= \label{I_1} - 
\frac{8}{\tau} \int \frac{d\w}{2\pi} K_1^{\alpha\beta}(\omega ) 
\langle f(\epsilon-\omega, \vec n)\rangle_n.   
\eeqa
\end{mathletters}
One can easily see that the elastic collision integral vanishes in
equilibrium $f(\epsilon, \vec n)= f(\epsilon),\ E=0$.  The explicit
definitions of $K_0$, $K_1$, and $L_0$ entering Eqs.~(\ref{I's}) are:
\begin{mathletters}
\label{kernels}
\beqa
&&K_1^{\alpha\beta}(\omega ) =
\Im \int \frac{d^2q}{(2\pi)^2}  {\cal{D}}^R(\omega,\vec q) 
\label{K1}
\\
&&\quad \times\left\{ \< n_\alpha D \> \< D n_\beta \> -  
\frac{\delta_{\alpha \beta}}{2} \left( \< D \> \< D \> + 
i \frac{\partial}{\partial \omega} \< D \> \right) \right\},
\nonumber
\\
&&
K_0^{\alpha\beta}(\omega )  = 
\Im \int \frac{d^2q}{(2\pi)^2} {\cal{D}}^R(\omega,\vec q)
\label{K0}\\
&&
\quad \times 
\left\{ \< n_\alpha D n_\beta \>\<  D \>
 -\frac{i}{v_F}\frac{\partial}{\partial q_\alpha}  \< D n_\beta \>
- \< D n_\alpha \>\< D n_\beta \>
\right\}, 
\nonumber\\
&&
L_0^{\alpha\beta}(\omega) =
- \Re \int \frac{d^2q}{(2\pi)^2} {\cal{D}}^R(\omega,\vec q) 
\label{L0} \\
&&
\quad\times 
\left\{\<  D \>  \frac{\partial}{\partial q_\beta} \< n_\alpha D \>
-\< D n_\alpha \>\frac{\partial}{\partial q_\beta}\<  D \>
- \< D n_\alpha \frac{\partial}{\partial q_\beta} D \>\right\}.
\nonumber
\eeqa
\end{mathletters}
Here, ${\cal D}^R$ denotes the retarded interaction propagator in singlet
channel \cite{intermed} and
 the angular averaging is
\begin{eqnarray*}
\<a  D b\> &\equiv& \int\frac{d \theta d\theta'}{(2\pi)^2}
a(\vec n)
D(\vec n, \vec n'; \omega, \q )b(\vec n'), 
\\ 
&&
\nonumber\\
\<a  D b D c\> &\equiv& \int\frac{d \theta d\theta'd\theta''}{(2\pi)^3}
a(\vec n)
D(\vec n, \vec n')b(\vec n')D(\vec n', \vec n'')c(\vec n'')
\end{eqnarray*}
\noindent
for arbitrary functions $a,\ b$.
The function $D(\vec n, \vec n'; \omega, \q )$ 
describes the classical motion of a particle on the
energy shell $\epsilon_F$ in a magnetic field:
\beqa
&&\left[-i\omega + i v_F \vec n \vec q +  
\omega_c\vec b
\left(\vec n \times 
\frac{\partial}{\partial \vec n} \right)\right] 
D(\vec n, \vec n') \label{Diffuson} \\ 
&&
+\frac{1}{\tau} \left[ D(\vec n, \vec n')
 - \langle  D(\vec n, \vec n') 
 \rangle_n \right] 
 = 2\pi \delta(\theta - \theta'). \nonumber 
\eeqa

It is noteworthy, that unlike in the ordinary Boltzmann equation,
the Lorentz force affects also the collision processes as one can see from
Eqs.~(\ref{kernels}) and (\ref{Diffuson}).

The above equations are written for the interaction in the singlet
channel only. In a situation where both triplet and singlet channels
are present, but the distribution function does not have a spin
structure (no Zeeman splitting or non-equilibrium spin occupation
present), one has to replace in Eq.~(\ref{kernels})
\beq
{\cal D}^R \to {\cal D}^R_s + 3 {\cal D}^R_t.
\label{r1}
\eeq
\noindent

{\em Calculation of the Hall coefficient} --
We calculate the Hall resistivity $\rho_{xy}$ 
in the first order in magnetic field.
We notice that due to the rotational and reflection
symmetries of the system 
\beq
K^{\alpha\beta}_i(\w ) = \delta_{\alpha\beta}K_i(\w) + 
\epsilon_{\alpha\beta}\w_c \tau \tilde{K}_i(\w),
\eeq 
and the same structure for the $L$ - kernel.
Here $\epsilon_{\alpha\beta}$ is the antisymmetric tensor, $\epsilon_{xy}=1$.

In order to calculate the conductivity we look for a solution of
Eqs.~(\ref{Boltzmann}) - (\ref{I's}) 

\beq
f(\vec n,\epsilon) = f_F(\epsilon) + { n}_\alpha \Gamma^{(1)}_\alpha(\epsilon) +
(\w_c\tau) {n}_\alpha \epsilon_{\alpha\beta}  \Gamma^{(2)}_\beta(\epsilon),
\label{Ansatz}
\eeq 

\noindent
where $f_F(\epsilon)=1/(e^{\epsilon/T}+1)$ is the Fermi distribution
function (all the energies are counted from the Fermi level), and the
unknown quantities ${\Gamma^{(1,2)}}$ are first order
in the electric field and zeroth order in the magnetic field. 

Substituting Eq.~(\ref{Ansatz})  into Eqs.~(\ref{Boltzmann}-\ref{I's}), 
we obtain the following system of equations:
\end{multicols}
\widetext
\begin{mathletters}
\beqa \label{oldstory}
e v_F E_\alpha \frac{\partial f_F(\epsilon) }{\partial \epsilon}
= &-& \frac{ \Gamma^{(1)}_\alpha (\epsilon)}{\tau} 
- \frac{4}{\tau}\int\frac{d\w}{2\pi}
\left[ K_1(\w) 
f_F(\epsilon-\w)\Gamma^{(1)}_\alpha (\epsilon) +
K_0(\w) f_F(\epsilon)\Gamma^{(1)}_\alpha (\epsilon-\w)
\right] \\
&-& \frac{4f_F(\epsilon)}{\tau}\int\frac{d\w}{2\pi}
L_0(\w)
e E_\alpha\frac{\partial}{\partial \epsilon} 
f_F(\epsilon-\w),
\nonumber \\
\Gamma^{(1)}_\alpha(\epsilon) 
= &-& \Gamma^{(2)}_\alpha(\epsilon)
- {4}\int\frac{d\w}{2\pi} 
\label{hallstory} 
f_F(\epsilon-\w) 
\left[ K_1(\w) 
\Gamma^{(2)}_\alpha(\epsilon) + 
\tilde{K}_1(\w) \Gamma^{(1)}_\alpha(\epsilon) \right] \\
&-& {4}\int\frac{d\w}{2\pi} f_F(\epsilon) \nonumber
\left[ K_0(\w) 
\Gamma^{(2)}_\alpha(\epsilon-\w) +
\tilde{K}_0(\w) \Gamma^{(1)}_\alpha(\epsilon-\w) \right] 
- {4f_F(\epsilon)}\int\frac{d\w}{2\pi} 
\frac{\partial}{\partial \w_c}\tilde{L}_(\w) eE_{\alpha} 
\frac{\partial}{\partial \epsilon} 
f_F(\epsilon-\w). \nonumber
\eeqa
\end{mathletters}
We solve Eqs.~(\ref{oldstory}) by iteration and obtain 
\begin{mathletters}
\beqa \label{gamma1sol}
\Gamma^{(1)}_\alpha (\epsilon)
= &-&  e v_F\tau E_\alpha \frac{\partial f_F(\epsilon) }{\partial \epsilon} 
+ {4 e v_F\tau}\int\frac{d\w}{2\pi}
\left[ K_1(\w) 
f_F(\epsilon-\w)\frac{\partial f_F(\epsilon) }{\partial \epsilon} + 
\quad K_0(\w) 
f_F(\epsilon)\frac{\partial f_F(\epsilon-\w) }{\partial \epsilon}
\right] E_\alpha \\ 
&-& {4f_F(\epsilon)}\int\frac{d\w}{2\pi} L_0(\w) 
e E_\alpha \frac{\partial}{\partial \epsilon} f_F(\epsilon-\w),
\nonumber
\\
\Gamma^{(2)}_\alpha = &-& \Gamma^{(1)}_\alpha 
- 4 e v_F \tau \int\frac{d\w}{2\pi} f_F(\epsilon-\w) 
\frac{\partial f_F(\epsilon) }{\partial \epsilon}
\left[ K_1(\w)  -
\tilde{K}_1(\w) \right]
E_\alpha  \label{gamma2sol}\\
&-& 4 e v_F \tau \int\frac{d\w}{2\pi} f_F(\epsilon)
\frac{\partial f_F(\epsilon-\w) }{\partial \epsilon}
\left[ K_0(\w) - \tilde{K}_0(\w) \right] E_\alpha 
- {4f_F(\epsilon)} \int\frac{d\w}{2\pi} 
\tilde{L}_0(\w) e E_{\alpha} 
\frac{\partial}{\partial \epsilon} f_F(\epsilon-\w). 
\nonumber
\eeqa
\label{ga}
\end{mathletters}
We substitute Eq.~(\ref{ga}) into Eq.~(\ref{Ansatz}) 
[$\Gamma^{(1)}$ does not contribute to the Hall 
current], and the result
into Eq.~(\ref{current}). 
Performing angular averaging and integration over
$\epsilon$ one finds components of the conductivity tensor. 

Rather than writing explicit results for the Hall conductivity
we present the expression for the Hall resistivity,
$
\rho_{xy} = \rho_H^D + \delta \rho_{xy}, \quad 
\frac{\delta \rho_{xy}}{\rho_H^D} = - 
\frac{\delta \sigma_{xy}}{(\w_c \tau) \sigma_D} -
 2 \frac{\delta \sigma_{xx}}{\sigma_D},
$
where $\sigma_D=e^2\nu v_F^2\tau/2$ is the Drude conductivity,
and $\rho_H^D$ is the classical Hall resistivity 
($\rho_H^D = - \w_c\tau/\sigma_D$). We find
\beq
\label{rescor}
\frac{\delta \rho_{xy}}{\rho_{H}^D} = 
\int_{-\infty}^{\infty} \frac{d\w}{\pi}
\frac{\partial}{\partial \w} \left( \w \coth \frac{\w}{2T} \right)  
\left[ \tilde{K}_1(\w) - \tilde{K}_0(\w) + 
\frac{\tilde{L}_0(\w)}{v_F \tau} + \frac{L_0(\w)}{v_F\tau} \right].
\eeq
\noindent
What remains is to find the explicit expressions for the kernels
entering into Eq.~(\ref{rescor}).
To do so, we  solve 
Eq.~(\ref{Diffuson}) up to the first order in magnetic field: 
\begin{mathletters}\label{diff}
\beqa 
&& D(\nn,\nn') = 
D_1(\nn, \nn') - \w_c \int \frac{d\theta_1}{2\pi}
D_1(\nn, \nn_1) \left( \nn_1 \times \frac{\partial}{\partial \nn_1}\right)
D_1(\nn_1, \nn'), \label{diffh}\\
&&D_1(\nn,\nn') = 
2\pi{\delta(\theta - \theta')}
D_0(\nn) + D_0(\nn)D_0(\nn')
\frac{C}{C\tau-1}, \\
&& C = \sqrt{\left(-i\w + 1/\tau\right)^2 + v_F^2 q^2}, 
\quad \quad D_0(\nn)=\frac{1}{ -i\w + i v_F \nn \q+1/\tau}.
\nonumber
\eeqa
\end{mathletters}
\noindent
We note, that only the second term on the R.H.S. of Eq.~(\ref{diffh}) contributes 
to $\tilde{K}_i$ and $\tilde{L}_0$.
Substituting Eqs.~(\ref{diff}) into Eqs.~(\ref{kernels}), the resulting kernels
into Eq.~(\ref{rescor}) and performing the angular 
integration we obtain the results for the correction to 
the diagonal conductivity, $\delta \sigma_{xx}$, and to the Hall resistivity,
\beq
\left( 
\matrix{
{\delta \rho_{xy}}/{\rho_{H}^D} \cr
{\delta \sigma_{xx}}/{\sigma_{D}}
}\right)=
\Im \int_{-\infty}^{\infty} \frac{d\w}{\pi}
\frac{\partial}{\partial \w} \left( \w \coth \frac{\w}{2T} \right)
\int \frac{qdq}{4\pi} {\cal D}^R(\w, q)
\left( 
\matrix{
B_{xy}(\w, q)\cr
B_{xx}(\w, q)
}\right),
\label{matrixForm}
\eeq
where the form-factors are defined as
\begin{mathletters}
\label{bs}
\beqa \label{explicit}
B_{xy}
&=&
-\left\{
\frac{2 v_F^2 q^2 /\tau^2 }{C^3 (C - 1/\tau)^3} 
+ \frac{v_F^2 q^2 [2C-5(-i\w + 1/\tau)]}{2 \tau^2 C^5 (C-1/\tau)^2}
+ \frac{(-i\w + 1/\tau) [C-(-i\w + 1/\tau)]}{\tau^2 C^4 (C-1/\tau)^2}
\right\}, \label{bxy}\\
B_{xx}
&=&
\left\{
\frac{ v_F^2 q^2/\tau^2}{C^3 (C - 1/\tau)^3} 
+ 
\frac{3 v_F^2 q^2}{2 \tau C^3 (C-1/\tau)^2} 
+
\frac{2[C-(-i\w + 1/\tau)]}{C(C-1/\tau)^2} 
+
\frac{2C-1/\tau}{C v_F^2 q^2} \left( \frac{C-(-i\w + 1/\tau)}{C-1/\tau} \right)^2
\right\}. \label{bxx}
\eeqa
\end{mathletters}
Expression for the form-factor $B_{xx}$ was obtained before in
Ref.~\onlinecite{intermed} and cited here for comparison. It is noteworthy
that $B_{xy,xx}(\w, q=0)=0$ as it is dictated by the gauge invariance.

\begin{multicols}{2}

{  
\narrowtext  
\begin{figure}[ht]  
%\vspace{0.1 cm} 
\epsfxsize=5 cm  
{\epsfbox{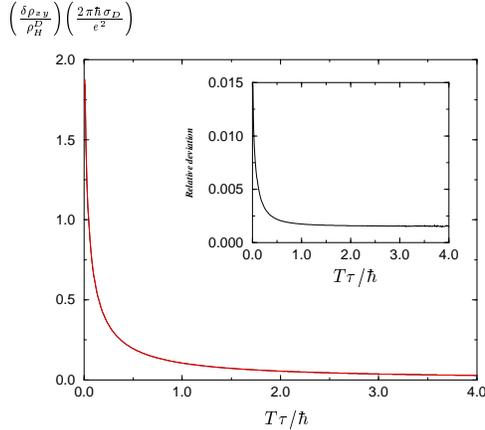}} 
%\vspace*{-6cm} 
\caption{Singlet channel correction to Hall resistivity 
$\delta \rho^{\rho}_{xy}$, given by Eq.¬(\ref{exactsinglet})
(solid line) 
and interpolation formula (\ref{inter1}) (dashed line).
On this scale two curves are indistinguishable, and 
the relative deviation is blown up on 
the inset.}  
\label{singletpic} 
\end{figure} 
} 

Equations (\ref{bs}) enable one to anticipate the behavior of parameter $\gamma$
from Eq.~(\ref{ourgamma}) even before the form of the interaction in
Eq.~(\ref{matrixForm}) is specified. At low temperatures, the integrals are 
determined by $\omega, \; qv_F \ll 1/\tau$. In this case both formfactors
in Eq.~(\ref{bs}) are controlled by the first terms in the R.H.S. As a result,
one  arrives to Eq.~(\ref{two}). At larger $q \gg 1/v_F\tau$, 
however, the
behavior of the formfactors is completely different, $B_{xx} \propto 1/q^2, \
B_{xy} \propto 1/q^4$, so the relation (\ref{two}) can no longer hold.

To study the details of temperature behavior of $\gamma$,
 one needs to evaluate 
the integrals in Eq.~(\ref{matrixForm}) for both the singlet and the 
triplet channel propagator (see Eq.~(50) and Eq.~(60) in 
Ref.~\onlinecite{intermed}):
\[ 
{\cal D}^R_t = - \frac{1}{\nu} 
\frac{C-1/\tau}{i\w + \frac{F_0^\sigma+1}{F_0^\sigma} (C-1/\tau)}.
\]
where $F_0^\sigma$ is the interaction constant in the triplet channel,
and  singlet propagator, ${\cal D}^R_s$, is obtained by 
putting $F_0^\sigma \to \infty$.
For completely spin-polarized electrons
only the  singlet channel correction is present. We find 
\beqa
&& \frac{\delta \rho^{\rho}_{xy}}{\rho_{H}^D} = - 
\frac{e^2}{4 \pi^2 \sigma_D} \left\{
- 4 {\bf C} + \frac{1}{24\pi T\tau} \psi' \left(1+\frac{1}{2\pi T \tau}\right) 
 \right. \nonumber \\
&& - \frac{11}{12} \psi \left(1+\frac{1}{2\pi T \tau}\right) - 
\frac{19}{2} \int_0^1 dx x \psi \left(1+\frac{x}{2\pi T \tau}\right) 
\nonumber \\
&& \left. +5 \int_0^1 dx x^2 \psi \left(1+\frac{x}{2\pi T \tau}\right) \right\}, 
\label{exactsinglet}
\eeqa
where ${\bf C} \approx 0.577$ is Euler's constant, $\psi(x)$ is the di-gamma
function. Equation~(\ref{exactsinglet}) 
can be approximated by the following interpolation formula
\beq \label{inter1}
\frac{\delta \rho^{\rho}_{xy}}{\rho_{H}^D} =
\frac{e^2}{\pi^2 \hbar \sigma_D} 
\ln \left(1 + \frac{11 \pi}{192} \frac{\hbar}{T\tau} \right).
\eeq
This formula reproduces the logarithmic behavior\cite{halls} 
in the diffusive limit ($T\tau \ll \hbar$) in accordance with 
Eq.~(\ref{two}), and
$
\frac{\delta \rho^{\rho}_{xy}}{\rho_{H}^D} = 
\frac{e^2}{\pi^2 \hbar \sigma_D} \frac{11 \pi}{192} \frac{\hbar}{T\tau}, 
$ for  $T\tau \gg \hbar$,
in  contrast with the linear dependence 
of the diagonal conductivity in this regime\cite{intermed}.
Finally, at intermediate temperatures Eq.~(\ref{inter1}) 
provides  accurate (for numerical reasons) description of the crossover, 
see Fig.~\ref{singletpic}.

The triplet channel correction is calculated analogously and 
the  interpolation formula similar to Eq.~(\ref{inter1}) reads
\beqa
\frac{\delta \rho^{\sigma}_{xy}}{\rho_{H}^D} = \frac{3e^2}{\pi^2 \hbar \sigma_D} 
&& \left\{1 - \frac{1}{F_0^\sigma} 
\ln \frac{1+F_0^\sigma + (T\tau/\hbar) g(F_0^\sigma)}{1 + T \tau/\hbar }
\right\} \times \nonumber \\
&& \quad \ln \left(1 + \frac{11 \pi}{192} \frac{\hbar}{T\tau} \right). 
\label{interp2}
\eeqa
Once again, Eq.~(\ref{interp2}) reproduces the asymptotic behavior 
at high and low temperature and gives numerically accurate results in the
intermediate region.
The definition of the smooth function $g(x)$ is 
\beqa
&& \ln g(x) = \frac{4}{11} \left[ - 5 f_3(x)- 12 f_2(x) - 
3 f_1(x) + 4 f_0(x) \right], \nonumber \\
&& \quad \quad f_j(x) = \frac{1}{x^{j}} 
\left( \ln [1+x] + \sum_{n=1}^{j} \frac{(-x)^n}{n} \right),
\eeqa
and it has the following asymptotic behavior,
$g(x) = 1 + x + \frac{3}{22} x^2 + \dots, \ x \ll 1$, and
$g(x) \to e^{-{70}/{33}},\ x \to -1$. The latter asymptotic is realized
when the system is close to the Stoner instability.
{  
\narrowtext  
\begin{figure}[ht]  
%\vspace{0.1 cm} 
\epsfxsize=5 cm  
{\epsfbox{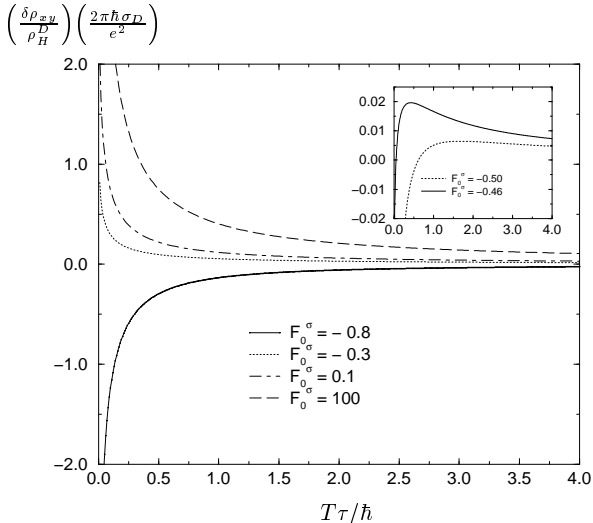}} 
%\vspace{0.2cm} 
\caption{Total correction to Hall resistivity 
$\delta \rho_{xy}$ for different values of the Fermi 
liquid parameter $F_0^\sigma$. For $-0.61 < F_0^\sigma < -0.45$,
the temperature dependence is non-monotonous, though very weak,
see inset. 
}
 
\label{fullpic} 
\end{figure} 
} 
The total correction to the Hall coefficient, $\delta \rho_{xy} = 
\delta \rho_{xy}^\rho + \delta \rho_{xy}^\sigma$, 
is shown on Fig.~\ref{fullpic}. Let us note in passing, that if the electron
system is polarized by strong in-plane field, only the singlet correction
(\ref{inter1}) remains.

Finally, we present the temperature dependence of the 
 quantity $\gamma$ from Eq.¬(\ref{ourgamma}). The plot shown 
 in Fig.~\ref{gammapic} was obtained with the help of Eqs.~(\ref{inter1}),
 (\ref{interp2}) of the present paper and Eq. (16) of 
 Ref.~\onlinecite{intermed}. We note, that  
the slope of this curve at zero temperature is always finite.
According to the figure, $\gamma(T)$'s 
deviation from the value of $2$ happens already 
at much lower temperatures than expected: it changes by a factor
of $2$ at about 
$T\tau /\hbar \simeq 0.1$ for the weak coupling ($F_0^\sigma \ll 1$) case. 

{  
\narrowtext  
\begin{figure}[ht]  
\vspace*{2.1 cm} 
\epsfxsize=4.6 cm  
{\epsfbox{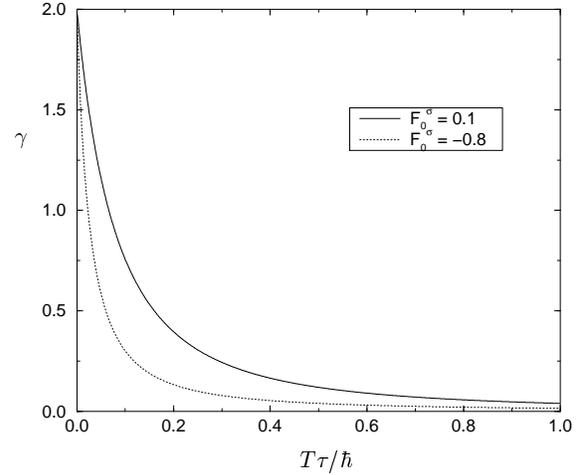}} 
\caption{Temperature dependence of the 
parameter $\gamma$, see Eq.~(\protect\ref{ourgamma}), relating the Hall 
coefficient to $\sigma_{xx}$.}  
\label{gammapic} 
\end{figure} 
} 
   
{\em Summary} -- To conclude, we investigated the temperature dependence
of the Hall resistivity of 2D electron gas 
for arbitrary values of $T\tau/\hbar$. It is shown that whereas the
relation between the Hall and diagonal resistivity (\ref{two}) indeed holds
for $T\tau \ll \hbar$,
it is rapidly violated at higher temperatures, see Fig.~\ref{gammapic}.  
  
{\em Acknowledgments} -- We are grateful to B.L. Altshuler, S.V. Kravchenko,
M.Yu. Reyzer, and A.K. Savchenko for stimulating discussions.
One of us (I.A.) was supported by the Packard foundation. Work in Lancaster
 University was partially funded by EPSRC-GR/R01767.

% **********************************************************************

\end{multicols}

\begin{references}

\bibitem{aar} B.L. Altshuler and A.G. Aronov in
{\em Electron-Electron Interactions in Disordered Systems}, 
eds. A.L. Efros, M. Pollak (North-Holland, Amsterdam, 1985).
\bibitem{halls} B.L. Altshuler {\em et. al.}, 
%D.E. Khmelnitskii, A.I. Larkin and P.A. Lee, 
Phys. Rev. B {\bf 22}, 5142 (1980). 
\bibitem{Kravchenko} For a recent review and references see
E. Abrahams, S. V. Kravchenko, and M. P. Sarachik,
Rev. Mod. Phys. {\bf 73}, 251 (2001).
\bibitem{Savchenko} Y.Y. Proskuryakov {\em et.al.}, 
Phys. Rev. Lett. {\bf 86}, 4895 (2001). 
\bibitem{intermed} G. Zala, B.N. Narozhny, and I.L. Aleiner, 
cond-mat/0105406.

\end{references}
\end{document}